\documentclass[RNAAS]{aastex62}
\usepackage{CJK}

\received{--}
\revised{--}
\accepted{--}

\shorttitle{A Preliminary Analysis of the Shangri-La Bolide on 2017 Oct 4}
\shortauthors{Ye}

\begin{document}
\begin{CJK*}{UTF8}{gbsn}

\title{A Preliminary Analysis of the Shangri-La Bolide on 2017 Oct 4}

\correspondingauthor{Quan-Zhi Ye}
\email{qye@caltech.edu}

\author[0000-0002-4838-7676]{Quan-Zhi Ye (叶泉志)}
\affiliation{Division of Physics, Mathematics and Astronomy, California Institute of Technology, Pasadena, CA 91125, U.S.A.}
\affiliation{Infrared Processing and Analysis Center, California Institute of Technology, Pasadena, CA 91125, U.S.A.}






\section*{}

At 12:07 UT (8:07 pm China Standard Time) on 2017 Oct 4, a bright bolide was widely observed in the Shangri-La region in the Province of Yunnan, China (Figure~1). The event was well observed by the general public as it took place on the night of the Mid Autumn Festival which associates with moon gazing. Sonic booms and ground shaking were reported in an area about a thousand square kilometers wide to the northwest of the Shangri-La City. Data from the U.S. government sensor suggested that the impact energy of the event is approximately 0.54 kt TNT equivalent, with the terminus of the bolide positioned at $28.1^\circ$~N, $99.4^\circ$~E. This is the largest observed bolide event overland since the bolide event took place in Mauritania on 2016 Jun 27 (1.2~kt).

\begin{figure}
\plotone{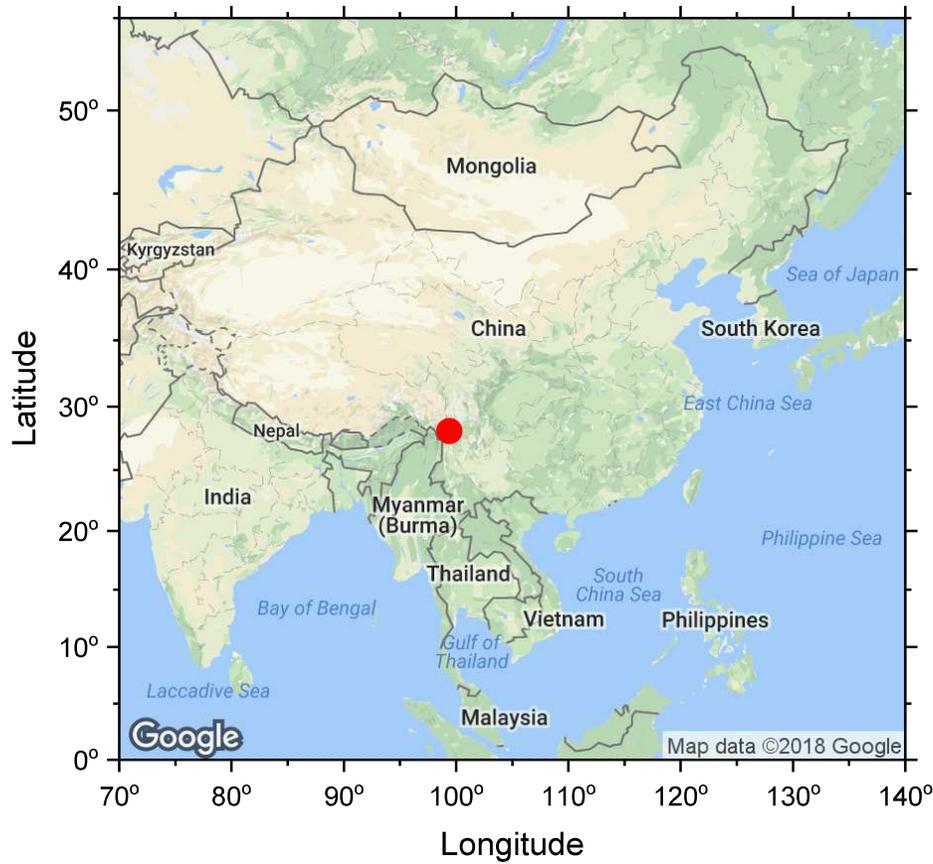}
\caption{Location of the Shangri-La bolide.}
\end{figure}

International Meteor Organization and American Meteor Society operates fireball report programs that collect world-wide observations of fireballs and bolides \citep[\url{https://www.imo.net/observations/fireballs/fireballs/} and \url{https://www.amsmeteors.org/fireballs/}, see][]{Hankey2014,Hankey2015}. However, very few (2) reports of the Shangri-La event have been archived in either database. Here we collect various of accounts from Weibo users, forum posts, and media reports, tabulated in Table~1.

\begin{table}[h!]
\tiny
\centering
\caption{Compilation of the visual and video accounts of the Shangri-La bolide. Some accounts are extracted from the post by Liu Wenjie\footnote{\url{http://bbs.tianya.cn/post-travel-821029-1.shtml}, in Chinese. Retrieved 2017 Dec 19.}.}
\begin{tabular}{lccccll}
\hline
\hline
Site & Coordinate & Concurrent & Delayed & Shaking? & Source & Note \\
& & boom? & boom? & & & \\
\hline
Balagezong National Park & $28.285^\circ$~N, $99.432^\circ$~E & \checkmark & \checkmark & \checkmark & Media report & ``Passing overhead and heading E, red'' \\
(巴拉格宗国家公园) & & & & & & \\
Baoshan City (保山) & $25.1^\circ$~N, $99.2^\circ$~E & & & & Media report & Multiple accounts \\
Benzilan (奔子栏) & $28.24^\circ$~N, $99.30^\circ$~E & \checkmark & \checkmark & \checkmark & Witness report & ``Sonic boom and shaking like a landslide is taking place'' \\
Dali City (大理) & $25.6^\circ$~N, $100.3^\circ$~E & & & & Media report & Multiple accounts \\
.. & $25.535^\circ$~N, $100.330^\circ$~E & & & & Dashcam & G56 freeway facing NW \\
Hongpo Village (红坡村) & $27.81^\circ$~N, $99.81^\circ$~E & \checkmark & \checkmark & & Media report & \\
Lijiang City (丽江) & $25.964^\circ$~N, $100.157^\circ$~E & & & & Dashcam & G5611 freeway facing NW \\
Niding Village (尼丁村) & $28.19^\circ$~N, $99.24^\circ$~E & \checkmark & \checkmark & \checkmark & Liu's post & House shaking, like a gas explosion \\
Nixi Township (尼西县城) & $28.1^\circ$~N, $99.5^\circ$~E & \checkmark & \checkmark & \checkmark & Media report & ``Pigs run out of barn'' \\
Nujiang City (怒江) & $25.8^\circ$~N, $98.9^\circ$~E & & & & Media report & Multiple accounts \\
Ramwok, Tibet (然乌) & $29.317^\circ$~N, $96.990^\circ$~E & & & & Dashcam & S201 highway facing E \\
Shusong Village (书松村) & $28.27^\circ$~N, $99.19^\circ$~E & \checkmark & \checkmark & \checkmark & Liu's post & \\
Tangman Village (汤满村) & $28.02^\circ$~N, $99.49^\circ$~E & \checkmark & \checkmark & \checkmark & Media report & \\
Xiaruo Township (霞若乡) & $27.80^\circ$~N, $99.30^\circ$~E & \checkmark & \checkmark & & Liu's post & \\
Xingfu Village (幸福村) & $28.14^\circ$~N, $99.43^\circ$~E & \checkmark & \checkmark & & Liu's post & \\
Shangri-La City & $27.8^\circ$~N, $99.7^\circ$~E & \checkmark & \checkmark & & Witness report & \\
(Zhongdian/中甸) & & & & & & \\
\hline
\end{tabular}
\end{table}

Figure~2 shows that moderate ground shaking was limited to an area about $30\times30~\mathrm{km^2}$ around Hongpo Village, $\sim40$~km northwest of the Shangri-La City (labeled as Deqen/迪庆藏族自治州 in the figure). Curiously, reports from Hongpo Village did not indicate ground shaking. It is not clear whether this was the case or the information was simply omitted. According to the local government agency, no significant property damage or loss has so far been reported\footnote{\url{http://news.chinaxiaokang.com/dujiazhuangao/2017/1006/260312.html}, in Chinese. Retrieved 2017 Dec 19.}.

We also collected 3 dashboard camera footages that the shooting locations are precisely known: two from freeways near the cities of Lijiang and Dali in Yunnan, approximately 100 and 300~km to the south, one from Rawu, Tibet, $\sim$300~km to the northwest (Figure~3). Initial analysis shows that the trajectory and motion of the bolide broadly agree the U.S. government sensor data. The Rawu footage included the full moon in the field of view, which appeared to be much less bright than the bolide at its peak brightness (Figure~3, frame b).

\begin{figure}
\plotone{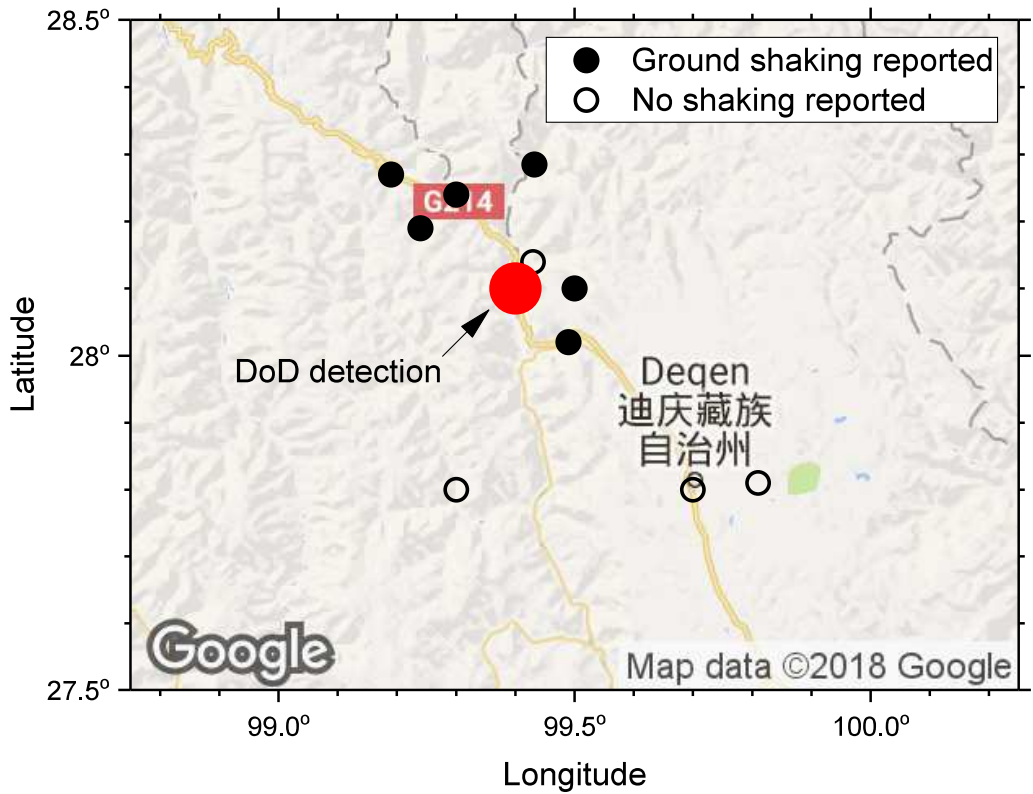}
\caption{Bolide witness and ground shaking reports in the Shangri-La area. Shangri-La City is labeled as Deqen/迪庆藏族自治州 in the figure. Also marked is the fireball detection reported by the U.S. Department of Defense (DoD) sensor.}
\end{figure}

\begin{figure}
\plotone{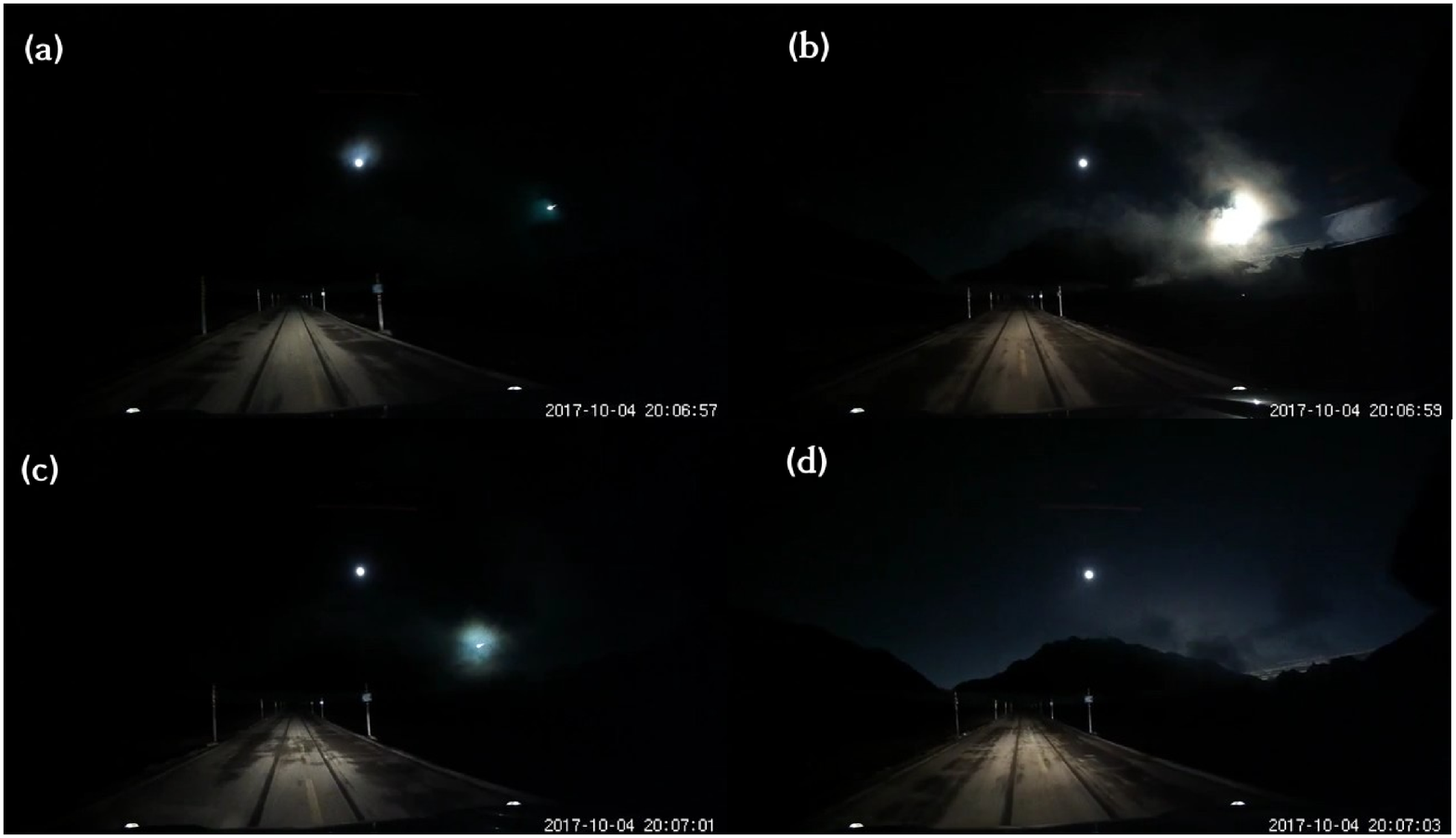}
\caption{Footage of the dashboard camera provided by Tan Kaixin, who was then traveling on an eastbound vehicle on Highway S201 near Rawu, Tibet. The clock of the camera was not calibrated and the timings on the lower-right corner are likely inaccurate.}
\end{figure}

The event was also detected seismically and infrasonically. China Earthquake Agency has reported positive detection by two seismic stations, Zhongdian ($27.824^\circ$~N, $99.707^\circ$~E) and Gongshan ($27.740^\circ$~N, $98.665^\circ$~E), registered at a Richter scale of 2.1\footnote{\url{http://news.xinhuanet.com/tech/2017-10/06/c_1121765328.htm}, in Chinese. Retrieved 2017 Dec 19.}. The CTBTO Kunming station and an infrasound array deployed in western Yunnan also detected the event (W. Su, private communication). Both datasets are not yet publicly available.

After the termination of ablation, meteorites decelerate to a speed of a few km/s and will typically spend a few minutes airborne before reaching the ground \citep{Ceplecha1998}. Atmospheric sounding data obtained by the nearby Xichang Station in Sichuan, $\sim$250~km to the east\footnote{Available from an archive maintained by the Department of Atmospheric Science at the University of Wyoming, \url{http://weather.uwyo.edu/upperair/sounding.html}, retrieved 2018 Jan. 4.}, revealed that the local atmosphere was dominated by westerly wind, largely parallel to the travel direction of the bolide. Therefore, we expect that any surviving meteorites would show little transverse motion. We estimate that the possible strewn field to be $\sim100$~km northeast of the bolide terminus near the Yunnan--Sichuan border (Figure~4). This is in line with independent calculation carried out by R. Matson (R. Matson, private communication).

We also attempted to use the Doppler weather radar data to look for micrometeorites deposits in the atmosphere. However, the predicted region is not covered by the Chinese weather radar network; the closest Doppler radar at Lijiang is about 200~km to the south with an effective observing range of 150~km.

\begin{figure}
\plotone{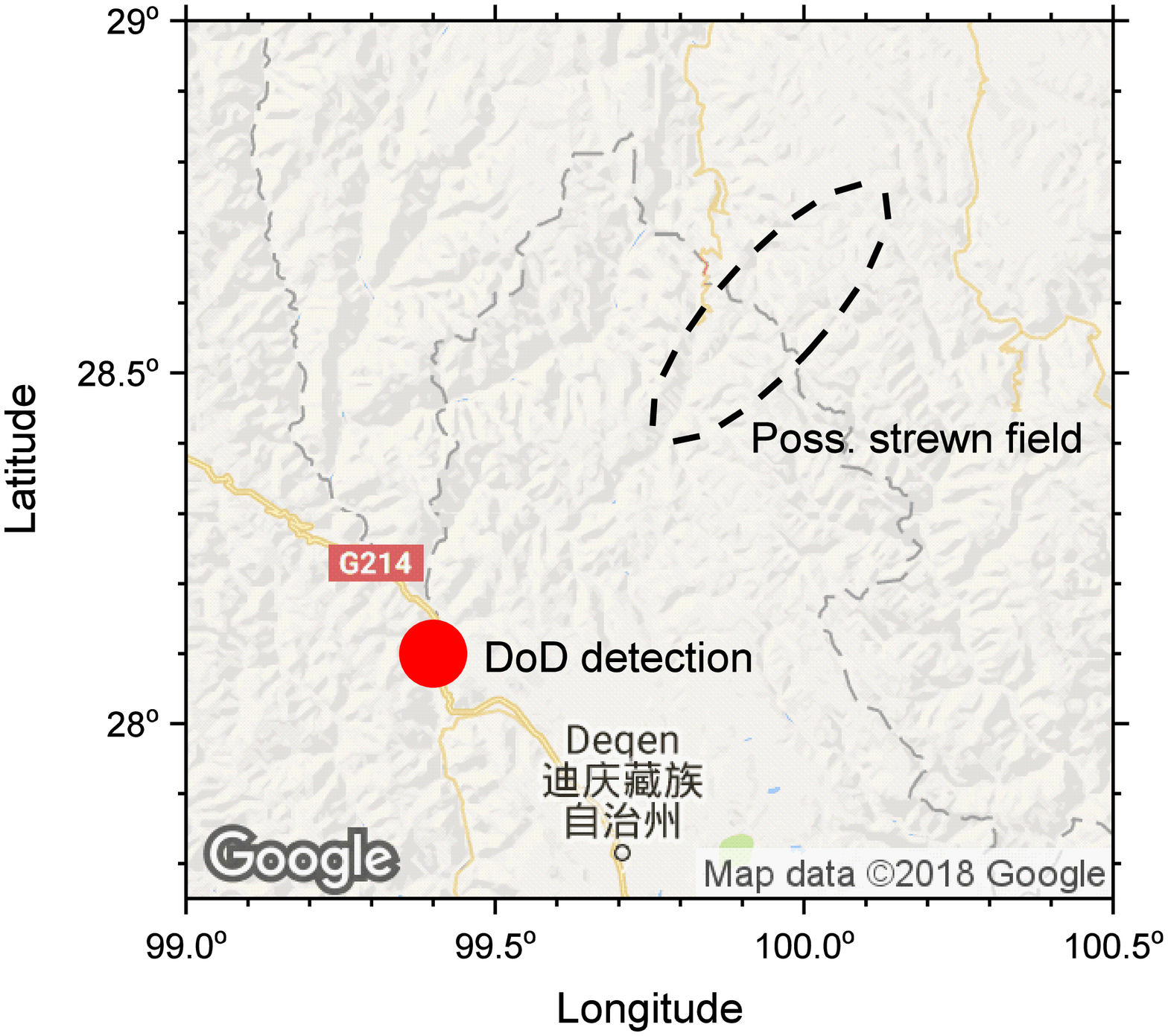}
\caption{Estimated strewn field of the Shangri-La meteorites.}
\end{figure}

It was reported that ``several hundreds of'' meteorite hunters had arrived on the scene to search for meteorites\footnote{\url{http://news.sina.com.cn/o/2017-10-10/doc-ifymrcmm9945166.shtml}, in Chinese. Retrieved 2018 Jan 3.}. Many chose to search in the northwest of Shangri-La City where most of the shock wave reports surfaced. Both this region and our estimated strewn field is very mountainous, but the latter is more so and is sparsely populated. At the time of this writing, the searches are ceased\footnote{\url{http://news.sina.com.cn/s/wh/2017-10-22/doc-ifymyyxw4058501.shtml}, in Chinese. Retrieved 2018 Jan 3.} and there is no credible successful recovery of the Shangri-La meteorite.

Unlike the U.S. and Europe, China is yet to establish a dedicated government-level body that oversees the research of near-Earth objects. Bolides that are energetic enough to be recorded by the U.S. government sensor occur in the country about once per year, with the last ones being the 2014 Xilin Gol event and the 2015 Gansu event. The Xilin Gol event, took place in the early morning of 2014 Nov 5 over the Gobi desert $\sim300$~km north of Beijing, was of similar magnitude to the Shangri-La event and had also caught considerable public attention. It was believed that multi-kilogram meteorites from the Xilin Gol event had reached the ground. Efforts (mostly led by amateurs) were launched to recover these meteorites but no successful recovery has been reported so far.

In a global scale, meteorite-dropping events are quite frequent. Using the event rate derived by \citet{Halliday1984} we estimate that event with $>1$~kg meteorite fall occurs once per week on average over the entire Chinese territory. Amateur astronomers in several parts of China have recently begun to build video camera networks, in the hope to better constrain the trajectory of the next meteorite-dropping events. As of early 2018, camera networks in Beijing, Shandong, Guangdong, northern Xinjiang and northwestern Tibet are operational \citep{Ye2018}.

\acknowledgments

I thank Peter Brown and Rob Matson for helpful discussions, as well as all the witnesses for sharing their videos and accounts on the internet. The author is supported by the GROWTH project (National Science Foundation Grant No. 1545949).

\end{CJK*}
\bibliographystyle{aasjournal}
\bibliography{ms}

\begin{thebibliography}{}
\expandafter\ifx\csname natexlab\endcsname\relax\def\natexlab#1{#1}\fi
\providecommand{\url}[1]{\href{#1}{#1}}
\providecommand{\dodoi}[1]{doi:~\href{http://doi.org/#1}{\nolinkurl{#1}}}
\providecommand{\doeprint}[1]{\href{http://ascl.net/#1}{\nolinkurl{http://ascl.net/#1}}}
\providecommand{\doarXiv}[1]{\href{https://arxiv.org/abs/#1}{\nolinkurl{https://arxiv.org/abs/#1}}}

\bibitem[{Ceplecha {et~al.}(1998)Ceplecha, Borovi{\v c}ka, Elford, Revelle,
  Hawkes, Porub{\v c}an, \& {\v S}imek}]{Ceplecha1998}
Ceplecha, Z., Borovi{\v c}ka, J., Elford, W.~G., {et~al.} 1998, \ssr, 84, 327,
  \dodoi{10.1023/A:1005069928850}

\bibitem[{Halliday {et~al.}(1984)Halliday, Blackwell, \&
  Griffin}]{Halliday1984}
Halliday, I., Blackwell, A.~T., \& Griffin, A.~A. 1984, Science, 223, 1405,
  \dodoi{10.1126/science.223.4643.1405}

\bibitem[{Hankey \& Perlerin(2015)}]{Hankey2015}
Hankey, M., \& Perlerin, V. 2015, in International Meteor Conference
  Mistelbach, Austria, ed. J.-L. {Rault} \& P.~{Roggemans}, 192--196

\bibitem[{Hankey {et~al.}(2014)Hankey, Perlerin, Lunsford, \&
  Meisel}]{Hankey2014}
Hankey, M., Perlerin, V., Lunsford, R., \& Meisel, D. 2014, in Proceedings of
  the International Meteor Conference, Poznan, Poland, 22-25 August 2013, ed.
  M.~{Gyssens}, P.~{Roggemans}, \& P.~{Zoladek}, 115--119

\bibitem[{Ye(2018)}]{Ye2018}
Ye, Q.-Z. 2018, Astronomical Research \& Technology.
\newblock \doarXiv{1708.00139}

\end{thebibliography}



\end{document}